\documentclass[12pt]{article}
\usepackage{amsmath,amssymb}
\usepackage{booktabs}
\usepackage{mathrsfs}
\usepackage[dvipdfmx]{graphicx}
\usepackage{graphicx}
\usepackage{multirow}

\newcommand\pubnumber{SNSN-323-63}
\newcommand\pubdate{\today}

 \def\support{\,}

\def\Title#1{\begin{center} {\Large #1 } \end{center}}
\def\Author#1{\begin{center}{ \sc #1} \end{center}}
\def\Address#1{\begin{center}{ \it #1} \end{center}}

\newcommand\pubblock{\rightline{\begin{tabular}{l} \pubnumber\\
         \pubdate  \end{tabular}}}
\newenvironment{Abstract}{\begin{quotation}  }{\end{quotation}}
\newenvironment{Presented}{\begin{quotation} \begin{center} 
             PRESENTED AT\end{center}\bigskip 
      \begin{center}\begin{large}}{\end{large}\end{center} \end{quotation}}
\def\Acknowledgements{\bigskip  \bigskip \begin{center} \begin{large}
             \bf ACKNOWLEDGEMENTS \end{large}\end{center}}

\def\beq{\begin{equation}}
\def\eeq#1{\label{#1}\end{equation}}
\def\eeqn{\end{equation}}

\def\beqa{\begin{eqnarray}}
\def\eeqa#1{\label{#1}\end{eqnarray}}
\def\eeqan{\end{eqnarray}}

\let\bar=\overbar

\def\Dslash{\not{\hbox{\kern-4pt $D$}}}
\def\dslash{\not{\hbox{\kern-2pt $\del$}}}

\def\msb{{\bar{\ssstyle M \kern -1pt S}}}

\begin{document}
\begin{titlepage}
\pubblock

\vfill
\Title{Exotic mesons with hidden charm and bottom near thresholds}
\vfill
\Author{ Shunsuke Ohkoda\support}
\Author{Y.~Yamaguchi$^1$, S.~Yasui$^2$, K.~Sudoh$^3$, A.~Hosaka$^1$}
\Address{$^1$Research Center for Nuclear Physics (RCNP), 
Osaka University, Ibaraki, Osaka, 567-0047, Japan}
\Address{$^2$KEK Theory Center, Institute of Particle and Nuclear
Studies, High Energy Accelerator Research Organization, 1-1, Oho,
Ibaraki, 305-0801, Japan}
\Address{$^3$Nishogakusha University, 6-16, Sanbancho, Chiyoda,
Tokyo, 102-8336, Japan}
\vfill
\begin{Abstract}
We study  heavy hadron spectroscopy near heavy meson 
 thresholds. We employ heavy pseudoscalar meson P and
heavy vector meson $\mathrm{P}^{\ast}$ as effective degrees of freedom 
 and consider meson exchange potentials between them. 
All possible composite states which can be constructed
 from the P and $\mathrm{P}^{\ast}$ mesons are studied up to the total
 angular momentum $J \le 2$. We consider, as exotic states, isosinglet states with
 exotic $J^{PC}$ quantum numbers and isotriplet states.  We solve
 numerically the Schr\"odinger equation with channel-couplings for each
 state. We found $\mathrm{B}^{(\ast)}\mathrm{\bar{B}}^{(\ast)}$ molecule states for $I^G(J^{PC}) = 1^+(1^{+-})$ 
 correspond to the masses of 
 twin resonances $\mathrm{Z}_{\mathrm b}(10610)$ and
 $\mathrm{Z}_{\mathrm b}(10650)$.  We predict several
 possible $\mathrm{B}^{(\ast)}\bar{\mathrm{B}}^{(\ast)}$ bound and/or resonant states in other channels.
On the other hand, there are no $\mathrm{D}^{(\ast)}\mathrm{\bar{D}}^{(\ast)}$ bound and/or resonant states whose quantum numbers are exotic.
\end{Abstract}
\vfill
\begin{Presented}
The 5th International Workshop on Charm Physics \\
(Charm2012) \\
14-17 May 2012, Honolulu, Hawai'i 96822
\end{Presented}
\vfill
\end{titlepage}
\def\thefootnote{\fnsymbol{footnote}}
\setcounter{footnote}{0}

\section{Introduction}
Recently, the exotic twin resonances $Z_b(10610)$ and $Z_b(10650)$ are discovered in the 
processes $\Upsilon(5S) \rightarrow \pi \pi \Upsilon(nS) (n=1,2,3)$ and 
$\Upsilon(5S) \rightarrow \pi \pi h_{b}(kP) (k=1,2)$ by Belle group 
\cite{Collaboration:2011gja,Belle:2011aa}.
The reported masses and widths of the two resonances are
$M(\mathrm{Z}_{\mathrm{b}}(10610)) = 10607.2 \pm 2.0$~MeV, 
$\Gamma (\mathrm{Z}_{\mathrm{b}}(10610)) = 18.4 \pm 2.4$~MeV and
$M (\mathrm{Z}_{\mathrm{b}}(10650)) = 10652.2 \pm 1.5$~MeV,
$\Gamma (\mathrm{Z}_{\mathrm{b}}(10650))=(11.5 \pm 2.2)$~MeV.
These resonances have some interesting properties.
First of all, $Z_{b}$'s have exotic quantum numbers $I^G(J^P) = 1^+(1^+)$.
Since $Z_{b}$'s are isotriplet states, they need four quarks as minimal constituents.
So $Z_{b}$'s are ``genuinely'' exotic states.
Secondly, $Z_{b}$'s have exotic decay ratios.
In general, a decay process with $h_b$ should be suppressed in heavy quark mass limit,
because these processes require a heavy quark spin flip.
Nevertheless, the decay rates of $\Upsilon(5S) \rightarrow Z_b \pi \rightarrow \pi \pi \Upsilon(nS)$
are comparable to those of $\Upsilon(5S) \rightarrow Z_b \pi \rightarrow \pi \pi h_b$.
Thirdly, $Z_b$'s are ``exotic twin'' resonances.
Their mass splitting is only 45 MeV ($\Delta m_{Z_b} \sim 45$ MeV).
This scale is not typically found in usual quarkonium.
These facts strongly suggest that $Z_{b}$'s have a molecular type structures as noted 
in Ref.~\cite{Bondar:2011ev,Voloshin:2011qa}

We can naively expect the existence of molecular states in heavy quark sectors.
We have two reasons.
One is that the kinetic term of Hamiltonian is suppressed.
Because the reduced mass is larger in heavy mesons.
Second is that $\mathrm{P}$ and $\mathrm{P}^{\ast}$ are degenerate thanks to heavy quark symmetry 
because the interaction of heavy quark spin is suppressed in heavy quark sector.
This leads the effects of channel-couplings become larger.

In this paper we study $\mathrm{P}^{(\ast)} \bar{\mathrm{P}}^{(\ast)}$ molecular states
in terms of the potential model.
We completely take into account the degeneracy of $\mathrm{P}$ and  $\mathrm{P}^{\ast}$
due to the heavy quark symmetry, and fully consider channel couplings of $\mathrm{P}^{(\ast)}$ and
$\bar{\mathrm{P}}^{(\ast)}$.
And we consider not only bound states but also resonant states.

This paper is organized as follows. In Sec.~2, we introduce 
(i) the $\pi$ exchange potential and (ii) the $\pi \rho\, \omega$ potential 
between $\mathrm{P}^{(\ast)}$ and $\bar{\mathrm{P}}^{(\ast)}$ mesons.
To obtain the potentials, we respect the heavy quark symmetry for
the $\mathrm{P}^{(\ast)}\mathrm{P}^{(\ast)}\pi$,
$\mathrm{P}^{(\ast)}\mathrm{P}^{(\ast)}\rho$ and
$\mathrm{P}^{(\ast)}\mathrm{P}^{(\ast)}\omega$ vertices. 
We classify all the possible states composed by a pair of $P^{(\ast)}$ and $\bar{P}^{(\ast)}$ mesons
with exotic quantum numbers $I^G(J^{PC})$.
In Sec.~3, we solve numerically the Schr\"odinger equations with channel-couplings and discuss 
the bound and/or resonant states of the $B^{(\ast)}\bar{B}^{(\ast)}$ and $D^{(\ast)}\bar{D}^{(\ast)}$
systems.
In Sec.~4 is devoted to summary.
\section{Method}
We employ the effective Lagrangians based on heavy quark and chiral symmetries.
They give the interaction Lagrangians of $\pi PP^{\ast}$ and $\pi P^{\ast}P^{\ast}$ with 
\begin{eqnarray}
 {\cal L}_{\pi PP^*} &=& 
 2 \frac{g}{f_\pi}(P^\dagger_a P^\ast_{b\,\mu}+P^{\ast\,\dagger}_{a\,\mu}P_b)\partial^\mu\hat{\pi}_{ab}
 \, , 
 \label{eq:piPP*}\\
 {\cal L}_{\pi P^*P^*} &=& 
 2 i \frac{g}{f_\pi}\epsilon^{\alpha \beta \mu \nu} v_{\alpha}
 P^{\ast\,\dagger}_{a\,\beta}P^\ast_{b\,\mu}\partial_{\nu}
 \hat{\pi}_{ab}
 \, ,
\label{eq:piP*P*}
\end{eqnarray}
where $f_{\pi} = 135$ MeV is the pion decay constant.
The coupling constant $|g| = 0.59$ is determined with reference to the observed decay width 
$\Gamma = 96$ keV for $\mathrm{D}^{\ast} \rightarrow \mathrm{D}\pi$,
assuming that the charm quark is sufficiently heavy.
And the interaction Lagrangians of $vPP$, $vPP^{\ast}$ and $vP^{\ast}P^{\ast}$  ($v=\rho$, $\omega$) are given by
\begin{eqnarray}
 {\cal L}_{vPP} &=&
-\sqrt{2}\beta g_V P_b P^{\dagger}_a v \cdot \hat{\rho}_{ba}
 \, , 
\label{eq:vPP} \\
{\cal L}_{vPP^*} &=& 
-2\sqrt{2}\lambda g_V v_{\mu}\epsilon^{\mu \nu \alpha \beta}
\left(P^\dagger_a
P^\ast_{b\,\beta}-P^{\ast\,\dagger}_{a\,\beta}P_b\right)
\partial_\nu(\hat{\rho}_\alpha)_{ba}
\, ,
\label{eq:vPP*} \\
{\cal L}_{vP^*P^*} &=&
\sqrt{2} \beta g_V P^*_b P^{*\dagger}_a v \cdot \hat{\rho}_{ba}\nonumber \\
&&+i2 \sqrt{2}\lambda g_V
P^{\ast\,\dagger}_{a\,\mu}P^\ast_{b\,\nu}
(\partial^\mu(\hat{\rho}^\nu)_{ba}-\partial^\nu(\hat{\rho}^\mu)_{ba}) \, ,
\label{eq:vP*P*}
\end{eqnarray}
where $g_V = 5.8$ is the coupling constant for $\rho \rightarrow \pi\pi$ decay.
The coupling constants are fixed as $\beta =0.9$ and $\lambda =0.56$ GeV, which are determined by the radiative decays of
 $\mathrm{D}^{\ast}$ meson and semileptonic decays of $\mathrm{B}$ meson with vector meson 
dominance by following Ref.~\cite{Isola:2003fh}.
Due to the $G$-parity, the signs of vertices for $v\bar{P}\bar{P}$, $v\bar{P}\bar{P}^{\ast}$ and
$v\bar{P}^{\ast}\bar{P}^{\ast}$ are opposite to those of $vPP$, $vPP^*$
and $vP^* P^*$, respectively, for $v=\omega$, while they are the same  
for $v=\rho$. 

Nex, we classify all the possible quantum numbers
$I^{G}(J^{PC})$ with isospin $I$, $G$-parity, total angular momentum $J$, parity $P$ and charge conjugation $C$ for the states which can be composed by a pair of $\mathrm{P}^{(\ast)}$ and $\bar{\mathrm{P}}^{(\ast)}$ mesons.
The charge conjugation $C$ is defined for $I=0$
or $I_{z}=0$ components
for $I=1$, and is related to the $G$-parity by $G=(-1)^{I}C$.
In the present discussion, we restrict upper limit of the total angular
momentum as $J \le 2$, because too higher angular momentum
will be disfavored to form bound or resonant states.
To derive a potential between $P$ and $P^{\ast}$, 
The $\mathrm{P}^{(\ast)}\bar{\mathrm{P}}^{(\ast)}$ components 
in the wave functions for various $J^{PC}$ are 
listed in Table~\ref{tbl:classification}. 
We use the notation $^{2S+1}L_{J}$ to denote the total spin $S$ and
relative angular momentum $L$ of the two body states of $\mathrm{P}^{(\ast)}$ and $\bar{\mathrm{P}}^{(\ast)}$ mesons.
The $J^{PC} = 0^{+-}$ state cannot be generated by a combination of $\mathrm{P}^{(\ast)}$ and $\bar{\mathrm{P}}^{(\ast)}$ mesons.
For $I=0$, there are many $\mathrm{B}^{(\ast)}\bar{\mathrm{B}}^{(\ast)}$
states whose quantum number $J^{PC}$ are the same as those of the quarkonia as shown
in the third row of $I=0$. 
In the present study, however, we do not consider these states,
because we have not yet included  mixing terms between the quarkonia
and the $\mathrm{P}^{(\ast)}\bar{\mathrm{P}}^{(\ast)}$ states. 
This problem will be left as future works.
Therefore, for $I=0$, we consider only the exotic
quantum numbers $J^{PC}=0^{--}$, $1^{-+}$ and $2^{+-}$.
The states of $I=1$ are
clearly not accessible by quarkonia. 
We investigate all possible 
 $J^{PC}$ states listed in Table~\ref{tbl:classification}.

We obtain the potentials with channel-couplings for each quantum number $I^{G}(J^{PC})$ in terms of
the interaction Lagrangians.
For each state, the Hamiltonian is given as a sum of the kinetic energy and the  potential with channel-couplings in a form of a matrix as
\begin{eqnarray}
H_{J^{PC}} = K_{J^{PC}} + \sum_{i=\pi, \rho, \omega}V^{i}_{J^{PC}}.
\end{eqnarray}
Breaking of the heavy quark symmetry is taken into account by mass difference between $\mathrm{P}$ and $\mathrm{P}^{\ast}$ mesons in the kinetic term.
The explicit forms of the Hamiltonian for each $I^{G}(J^{PC})$ are presented in Ref.~\cite{Ohkoda:2011vj}.
For example, the $J^{PC} = 1^{+-}$ state has four components,
$\frac{1}{\sqrt{2}} \left(
\mathrm{P}\bar{\mathrm{P}}^{\ast}-\mathrm{P}^{\ast}\bar{\mathrm{P}}
\right) (^{3}S_{1})$, $\frac{1}{\sqrt{2}} \left(
\mathrm{P}\bar{\mathrm{P}}^{\ast}-\mathrm{P}^{\ast}\bar{\mathrm{P}}
\right) (^{3}D_{1})$,
$\mathrm{P}^{\ast}\bar{\mathrm{P}}^{\ast}(^{3}S_{1})$,
$\mathrm{P}^{\ast}\bar{\mathrm{P}}^{\ast}(^{3}D_{1})$ and hence it
gives a  potential in the form of $4\times 4$ matrix.


\begin{table}[htdp]
\caption{ Various components of the
 $\mathrm{P}^{(\ast)}\bar{\mathrm{P}}^{(\ast)}$ states for several
 $J^{PC}$ ($J \le 2$). The exotic quantum numbers which cannot be
 assigned to charmonia or bottomonia $\mathrm{Q}\bar{\mathrm{Q}}$ are indicated by
 $\surd$. The $0^{+-}$ state cannot be neither bottomonium nor
 $\mathrm{P}^{(\ast)}\bar{\mathrm{P}}^{(\ast)}$ states.
Examples of bottomonia are shown in non-exotic quantum numbers.}
\begin{center}
{\renewcommand\arraystretch{1.5}
\footnotesize
\begin{tabular}{|c|c|c|c|}
\hline
$J^{PC}$ & components & \multicolumn{2}{c|}{exoticness} \\ 
\cline{3-4}
                         &                     & $I=0$ & $I=1$ \\ 
\hline
\hline
$0^{+-}$ & ------ & $\surd$ & $\surd$ \\ 
\hline
$0^{++}$ & $\mathrm{P}\bar{\mathrm{P}}(^{1}S_{0})$, $\mathrm{P}^{\ast}\bar{\mathrm{P}}^{\ast}(^{1}S_{0})$, $\mathrm{P}^{\ast}\bar{\mathrm{P}}^{\ast}(^{5}D_{0})$ & $\chi_{\mathrm{b}0}$ & $\surd$ \\ 
\hline
$0^{--}$ & $\frac{1}{\sqrt{2}} \left( \mathrm{P}\bar{\mathrm{P}}^{\ast}+\mathrm{P}^{\ast}\bar{\mathrm{P}} \right)(^{3}P_{0})$ & $\surd$ & $\surd$  \\ 
\hline
$0^{-+}$ & $\frac{1}{\sqrt{2}} \left( \mathrm{P}\bar{\mathrm{P}}^{\ast}-\mathrm{P}^{\ast}\bar{\mathrm{P}} \right)(^{3}P_{0})$, $\mathrm{P}^{\ast}\bar{\mathrm{P}}^{\ast}(^{3}P_{0})$ & $\eta_{\mathrm b}$ & $\surd$ \\ 
\hline
$1^{+-}$ & $\frac{1}{\sqrt{2}} \left( \mathrm{P}\bar{\mathrm{P}}^{\ast}-\mathrm{P}^{\ast}\bar{\mathrm{P}} \right) (^{3}S_{1})$, $\frac{1}{\sqrt{2}} \left( \mathrm{P}\bar{\mathrm{P}}^{\ast}-\mathrm{P}^{\ast}\bar{\mathrm{P}} \right) (^{3}D_{1})$, $\mathrm{P}^{\ast}\bar{\mathrm{P}}^{\ast}(^{3}S_{1})$, $\mathrm{P}^{\ast}\bar{\mathrm{P}}^{\ast}(^{3}D_{1})$ & $\mathrm{h}_{\mathrm b}$ & $\surd$ \\ 
\hline
$1^{++}$ & $\frac{1}{\sqrt{2}} \left( \mathrm{P}\bar{\mathrm{P}}^{\ast}+\mathrm{P}^{\ast}\bar{\mathrm{P}} \right) (^{3}S_{1})$, $\frac{1}{\sqrt{2}} \left( \mathrm{P}\bar{\mathrm{P}}^{\ast}+\mathrm{P}^{\ast}\bar{\mathrm{P}} \right)(^{3}D_{1})$, $\mathrm{P}^{\ast}\bar{\mathrm{P}}^{\ast}(^{5}D_{1})$ & $\chi_{\mathrm{b}1}$ & $\surd$ \\ 
\hline
$1^{--}$ & $\mathrm{P}\bar{\mathrm{P}}(^{1}P_{1})$, $\frac{1}{\sqrt{2}} \left( \mathrm{P}\bar{\mathrm{P}}^{\ast}+\mathrm{P}^{\ast}\bar{\mathrm{P}} \right)(^{3}P_{1})$, $\mathrm{P}^{\ast}\bar{\mathrm{P}}^{\ast}(^{1}P_{1})$, $\mathrm{P}^{\ast}\bar{\mathrm{P}}^{\ast}(^{5}P_{1})$, $\mathrm{P}^{\ast}\bar{\mathrm{P}}^{\ast}(^{5}F_{1})$ & $\Upsilon$ & $\surd$ \\ 
\hline
$1^{-+}$ & $\frac{1}{\sqrt{2}} \left( \mathrm{P}\bar{\mathrm{P}}^{\ast}-\mathrm{P}^{\ast}\bar{\mathrm{P}} \right)(^{3}P_{1})$, $\mathrm{P}^{\ast}\bar{\mathrm{P}}^{\ast}(^{3}P_{1})$ & $\surd$ & $\surd$ \\ 
\hline
$2^{+-}$ & $\frac{1}{\sqrt{2}} \left( \mathrm{P}\bar{\mathrm{P}}^{\ast}-\mathrm{P}^{\ast}\bar{\mathrm{P}} \right)(^{3}D_{2})$, $\mathrm{P}^{\ast}\bar{\mathrm{P}}^{\ast}(^{3}D_{2})$ & $\surd$ & $\surd$ \\ 
\hline
$2^{++}$ & $\mathrm{P}\bar{\mathrm{P}}(^{1}D_{2})$, $\frac{1}{\sqrt{2}} \left( \mathrm{P}\bar{\mathrm{P}}^{\ast}+\mathrm{P}^{\ast}\bar{\mathrm{P}} \right)(^{3}D_{2})$, $\mathrm{P}^{\ast}\bar{\mathrm{P}}^{\ast}(^{1}D_{2})$, $\mathrm{P}^{\ast}\bar{\mathrm{P}}^{\ast}(^{5}S_{2})$, $\mathrm{P}^{\ast}\bar{\mathrm{P}}^{\ast}(^{5}D_{2})$, $\mathrm{P}^{\ast}\bar{\mathrm{P}}^{\ast}(^{5}G_{2})$ & $\chi_{\mathrm{b}2}$ & $\surd$ \\ 
\hline
$2^{-+}$ & $\frac{1}{\sqrt{2}} \left( \mathrm{P}\bar{\mathrm{P}}^{\ast}-\mathrm{P}^{\ast}\bar{\mathrm{P}} \right)(^{3}P_{2})$, $\frac{1}{\sqrt{2}} \left( \mathrm{P}\bar{\mathrm{P}}^{\ast}-\mathrm{P}^{\ast}\bar{\mathrm{P}} \right)(^{3}F_{2})$, $\mathrm{P}^{\ast}\bar{\mathrm{P}}^{\ast}(^{3}P_{2})$, $\mathrm{P}^{\ast}\bar{\mathrm{P}}^{\ast}(^{3}F_{2})$ & $\eta_{\mathrm{b}2}$ & $\surd$ \\ 
\hline
$2^{--}$ & $\frac{1}{\sqrt{2}} \left( \mathrm{P}\bar{\mathrm{P}}^{\ast}+\mathrm{P}^{\ast}\bar{\mathrm{P}} \right)(^{3}P_{2})$, $\frac{1}{\sqrt{2}} \left( \mathrm{P}\bar{\mathrm{P}}^{\ast}+\mathrm{P}^{\ast}\bar{\mathrm{P}} \right)(^{3}F_{2})$, $\mathrm{P}^{\ast}\bar{\mathrm{P}}^{\ast}(^{5}P_{2})$, $\mathrm{P}^{\ast}\bar{\mathrm{P}}^{\ast}(^{5}F_{2})$ & $\psi_{\mathrm{b}2}$ & $\surd$ \\ 
\hline
\end{tabular}
}
\end{center}
\label{tbl:classification}
\end{table}%

\section{Numerical results}
To obtain the solutions of the $\mathrm{P}^{(\ast)}\bar{\mathrm{P}}^{(\ast)}$  states,
we solve numerically the Schr\"odinger equations which are  
second-order differential equations with channel-couplings.
As numerics, the renormalized Numerov method developed in
Ref.~\cite{johnson} is adopted.
The resonant states are found from the phase shift $\delta$ as a function of the scattering energy $E$.
The resonance position $E_r$ is defined by an inflection point of the 
phase shift $\delta(E)$ and the resonance width by 
$\Gamma_r = 2/(d\delta /dE)_{E=E_r}$.
To check consistency of our method with others,
we also use the complex scaling method (CSM) \cite{CSM}. We obtain an agreement 
in results between the renormalized Nemerov method and the CSM.

\begin{figure}[htbp]
\includegraphics[width=13cm]{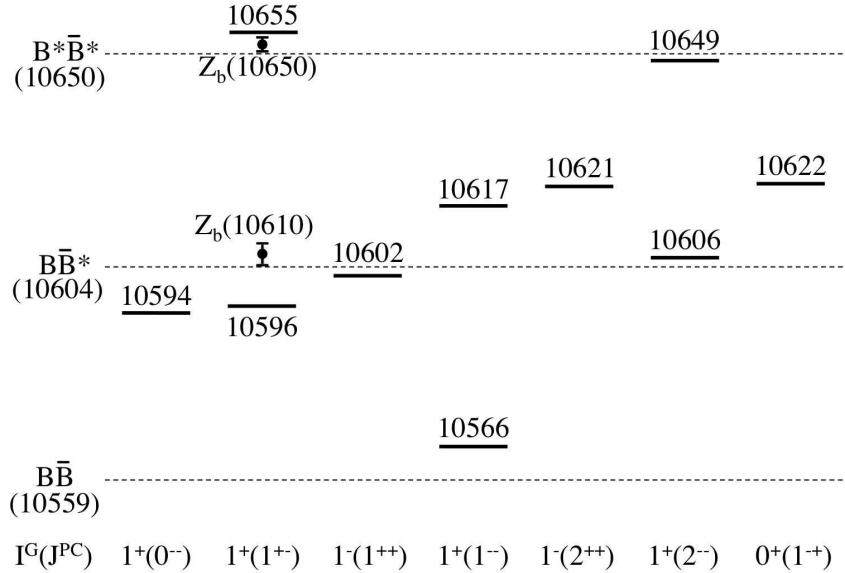}
\caption{The $\mathrm{B}^{(*)}\bar{\mathrm{B}}^{(*)}$ bound and
 resonant states with exotic $I^{G}(J^{PC})$. The dots with error bars denote the
 position of the experimentaly observed 
 $\mathrm{Z}_{\mathrm{b}}$'s where $M(\mathrm{Z}_{\mathrm{b}}(10610)) =
 10607.2$ MeV and $M (\mathrm{Z}_{\mathrm{b}}(10650)) = 10652.2$ MeV. Solid lines are for our predictions for the
 energies of the bound and resonant states when the $\pi \rho\, \omega$
 potential is employed. Mass values are shown in units of MeV.}
\label{fig:result}
\end{figure}

In Table~\ref{tbl:result_table}, we summarize the result of the obtained
 bound and resonant states, and their possible decay 
modes to quarkonium and light flavor meson.
For decay modes, the $\rho$ meson can be either real or virtual
 depending on the mass of the decaying particle, depending on the resonance energy which is either sufficient or not to emit the real state of $\rho$ or $\omega$ meson.
   $\rho^{\ast}(\omega^{\ast})$ indicates that it is a virtual state in radiative decays assuming the vector meson dominance.   
We show  the mass spectrum of these states in Fig~\ref{fig:result}.  

Let us see the states of isospin $I=1$.  
Interestingly, having the present potential we find the twin states in the $I^{G}(J^{PC})=1^{+}(1^{+-})$ near the
$\mathrm{B}\bar{\mathrm{B}}^{\ast}$ and  $\mathrm{B}^{\ast}\bar{\mathrm{B}}^{\ast}$
thresholds;  a bound state slightly below the $\mathrm{B}\bar{\mathrm{B}}^{\ast}$ threshold, and  a resonant state slightly above the $\mathrm{B}^{\ast}\bar{\mathrm{B}}^{\ast}$ threshold.
The binding energy is 8.5 MeV, and the resonance energy and decay width
are 50.4 MeV and 15.1 MeV, respectively, from the $\mathrm{B}\bar{\mathrm{B}}^{\ast}$ threshold.
The twin states are obtained when the $\pi \rho\, \omega$ potential is
used.  We interpret them as the
$\mathrm{Z}_\mathrm{b}(10610)$ and $\mathrm{Z}_\mathrm{b}(10650)$  observed in the Belle
experiment~\cite{Collaboration:2011gja,Belle:2011aa}.
It should be emphasized that the interaction in the present study has
been determined  in the previous works without knowing the experimental
data of $\mathrm{Z}_{\mathrm{b}}$'s.
 
Several comments are in order.  
First, the bound state of lower energy has been obtained in the coupled
channel method of $\mathrm{B}\bar{\mathrm{B}}^{\ast}$ and
$\mathrm{B}^{\ast}\bar{\mathrm{B}}^{\ast}$  channels.  
In reality, however, they also couple to other lower channels such
as $\pi h_b$,  $\pi \Upsilon$ and so on as shown in Table~\ref{tbl:classification}.
Once these decay channels are included, the bound state will be a resonant 
state with a finite width.  
A qualitative discussion will be given in Section~5.
Second, when the $\pi$ exchange potential is used, only the lower bound state is
obtained but the resonant state is not.  However, we have verified
that a small change in the $\pi$ exchange potential generates, as well as the
bound state, the corresponding resonant state also.   
Therefore, the pion dominance is working for the  $\mathrm{B}\bar{\mathrm{B}}^{\ast}$ and $\mathrm{B}^{\ast}\bar{\mathrm{B}}^{\ast}$ systems.
Third, it would provide a direct evidence of these states to be $\mathrm{B}\bar{\mathrm{B}}^{\ast}$ and $\mathrm{B}^{\ast}\bar{\mathrm{B}}^{\ast}$ molecules if the  $\mathrm{B}\bar{\mathrm{B}}^{\ast}$ and $\mathrm{B}^{\ast}\bar{\mathrm{B}}^{\ast}$ decays are observed in experiments.  
Whether the energies are below or above the thresholds is also checked by
the observation of these decays.  

In other channels, we  further predict the $\mathrm{B}^{(\ast)}\bar{\mathrm{B}}^{(\ast)}$ bound 
and resonant states.
The $I^{G}(J^{PC})=1^{-}(0^{++})$ state is a bound state with binding energy 6.5 MeV
from the $\mathrm{B}\bar{\mathrm{B}}$ threshold for the $\pi$ exchange potential, while no structure for the $\pi \rho\, \omega$ potential.
The existence of this state  depends on the
details of the potential, 
while the states in the other quantum numbers are rather robust.
Let us see the results for the latter states from the $\pi \rho\, \omega$ potentials. 
For $1^{+}(0^{--})$ and $1^{-}(1^{++})$, we find
bound states with binding energy 9.8 MeV and 1.9 MeV from the
$\mathrm{B}\bar{\mathrm{B}}^{\ast}$ threshold, respectively.
These bound states appear also for the $\pi$ exchange potential, though the binding energy of the $1^{-}(1^{++})$ state becomes larger.
The $1^{-}(2^{++})$ state is a resonant state with the resonance energy 62.7
MeV and the decay width 8.4 MeV.
The $1^{+}(1^{--})$ states are  twin resonances with the resonance energy 7.1
MeV and the decay width 37.4 MeV for the first resonance, and the
resonance energy 58.6 MeV and the decay width 27.7 MeV for the second.  
The resonance energies are measured from the $\mathrm{B}\bar{\mathrm{B}}$ threshold.
The $1^{+}(2^{--})$ states also form twin resonances with the resonance energy
2.0 MeV and the decay width 3.9 MeV for the first resonance and the
resonance energy 44.1 MeV and the decay width 2.8 MeV for the second, 
where the resonance energies have are  measured from the $\mathrm{B}\bar{\mathrm{B}}^{\ast}$ threshold.

Next we discuss the result for the states of isospin $I=0$.  
In general, the interaction in these states are either repulsive or 
only weakly attractive as compared to the cases of $I=1$.  
 The fact that there are less channel-couplings explains less attraction
 partly. 
Because of this, we find only one resonant state with $I^{G}(J^{PC})=0^{+}(1^{-+})$, 
as shown in Fig~\ref{fig:result} and in Table~\ref{tbl:result_table_2}.  
The $0^{+}(1^{-+})$ state is a resonant state with the resonance energy 17.8
MeV and the decay width 30.1 MeV for the $\pi \rho\, \omega$ potential.

In the present study, all the states appear in the threshold regions and
therefore  are all weakly bound or resonant states.
The present results are consequences of unique features of the bottom quark sector; the large reduced mass of the $\mathrm{B}^{(\ast)}\bar{\mathrm{B}}^{(\ast)}$ systems and the strong tensor force induced by the mixing of $\mathrm{B}$ and $\mathrm{B}^{\ast}$ with small mass splitting.
In fact, in the charm sector, our model does not predict any bound or
resonant states in the region where we research numerically. Because the reduced mass is smaller and the mass splitting between $\mathrm{D}$ and $\mathrm{D}^{\ast}$ is larger.  

\begin{table}[htbp]
\caption{\small Various properties of the
 $\mathrm{B}^{(\ast)}\bar{\mathrm{B}}^{(\ast)}$ bound and resonant
 states with possible $I^{G}(J^{PC})$ in $I=1$. The energies $E$ can be
 either pure real for bound states or complex for resonances.  The real
 parts are measured from the thresholds as indicated in the second
 column.  The imaginary parts are half of the decay widths of the
 resonances, $\Gamma/2$. In the last two columns, decay channels of a
 quarkonium and a light flavor meson are indicated. Asterisk of
 $\rho^{\ast}$ indicates that the decay occures only with a virtual
 $\rho$ while subsequently transit to a real photon via vector meson dominance.}  
\begin{center}
{
\small
\begin{tabular}{|c|c|c|c|c|c|}
\hline
$I^G (J^{PC})$ & threshold &  \multicolumn{2}{c|}{$E$ [MeV]} & \multicolumn{2}{c|}{decay channels} \\
\cline{3-6}
 & & $\pi$-potential & $\pi \rho\, \omega$-potential & s-wave & p-wave \\
\hline
$1^+ (0^{+-})$ & --- & --- & --- &  --- & $\mathrm{h}_{\mathrm{b}}+\pi$, $\chi_{\mathrm{b}0,1,2} \!+\! \rho$ \\
\hline
$1^- (0^{++})$ & $\mathrm{B}\bar{\mathrm{B}}$ & $-6.5$  & no & $\eta_{\mathrm{b}} \!+\! \pi$, $\Upsilon \!+\! \rho$ & $\mathrm{h}_{\mathrm{b}} \!+\! \rho^{\ast}$, $\chi_{\mathrm{b}1} \!+\! \pi$ \\
\hline
$1^+ (0^{--})$ & $\mathrm{B}\bar{\mathrm{B}}^{\ast}$ & $-9.9$ & $-9.8$ & $\chi_{\mathrm{b}1} \!+\! \rho^{\ast}$ & $\eta_{\mathrm{b}} \!+\! \rho$, $\Upsilon \!+\! \pi$ \\
\hline
$1^- (0^{-+})$ & $\mathrm{B}\bar{\mathrm{B}}^{\ast}$ & no & no & $\mathrm{h}_{\mathrm{b}} \!+\! \rho$, $\chi_{\mathrm{b}0} \!+\! \pi$ & $\Upsilon \!+\! \rho$ \\
\hline
\multirow{2}*{$1^+ (1^{+-})$} & \multirow{2}*{$\mathrm{B}\bar{\mathrm{B}}^{\ast}$} & 
 \multirow{2}*{$-7.7$} & $-8.5$  &\multirow{2}*{ $\Upsilon \!+\! \pi$} &
 \multirow{2}*{$\mathrm{h}_{\mathrm{b}} \!+\! \pi$, $\chi_{\mathrm{b}1}
 \!+\! \rho^{\ast}$}  \\
 & &  & $50.4-i15.1/2$ & &\\
\hline
$1^- (1^{++})$ & $\mathrm{B}\bar{\mathrm{B}}^{\ast}$ & $-16.7$  & $-1.9$ & $\Upsilon \!+\! \rho$ & $\mathrm{h}_{\mathrm{b}} \!+\! \rho^{\ast}$, $\chi_{\mathrm{b}0,1} \!+\! \pi$ \\
\hline
\multirow{2}*{$1^+ (1^{--})$} & \multirow{2}*{$\mathrm{B}\bar{\mathrm{B}}$} & 
 $7.0-i37.9/2$ & $7.1-i37.4/2$ &\multirow{2}*{ $\mathrm{h}_{\mathrm{b}} \!+\! \pi$,
 $\chi_{\mathrm{b}0,1,2} \!+\! \rho^{\ast}$} & \multirow{2}*{
 $\eta_{\mathrm{b}} \!+\! \rho$, $\Upsilon \!+\! \pi$} \\
 & & $58.8-i30.0/2$ & $58.6-i27.7/2$ & &\\
\hline
$1^- (1^{-+})$ & $\mathrm{B}\bar{\mathrm{B}}^{\ast}$ & no  & no & $\mathrm{h}_{\mathrm{b}} \!+\! \rho$, $\chi_{\mathrm{b}1} \!+\! \pi$ & $\eta_{\mathrm{b}} \!+\! \pi$, $\Upsilon \!+\! \rho$ \\
\hline
$1^+ (2^{+-})$ & $\mathrm{B}\bar{\mathrm{B}}^{\ast}$ & no & no & --- & $\mathrm{h}_{\mathrm{b}} \!+\! \pi$, $\chi_{\mathrm{b}0,1,2} \!+\! \rho$ \\
\hline
$1^- (2^{++})$ & $\mathrm{B}\bar{\mathrm{B}}$ & $63.5-i8.3/2$ & $62.7-i8.4/2$ & $\Upsilon \!+\! \rho$ & $\mathrm{h}_{\mathrm{b}} \!+\! \rho^{\ast}$, $\chi_{\mathrm{b}1,2} \!+\! \pi$ \\
\hline
$1^- (2^{-+})$ & $\mathrm{B}\bar{\mathrm{B}}^{\ast}$ & no & no & $\mathrm{h}_{\mathrm{b}} \!+\! \rho$ & $\Upsilon \!+\! \rho$ \\
\hline
\multirow{2}*{$1^+ (2^{--})$} & \multirow{2}*{$\mathrm{B}\bar{\mathrm{B}}^{\ast}$} & 
 $2.0-i4.1/2$ & $2.0-i3.9/2$ &\multirow{2}*{$\chi_{\mathrm{b}1} \!+\! \rho^{\ast}$} &
 \multirow{2}*{ $\eta_{\mathrm{b}} \!+\! \rho$, $\Upsilon \!+\! \pi$}  \\
 & & $44.2-i2.5/2$ & $44.1-i2.8/2$ & & \\
\hline
\end{tabular}
}
\end{center}
\label{tbl:result_table}
\end{table}%

\begin{table}[htbp]
\caption{\small The $\mathrm{B}^{(\ast)}\bar{\mathrm{B}}^{(\ast)}$ bound and resonant states with exotic $I^{G}(J^{PC})$ in $I=0$. (Same convention as Table II.)}
\begin{center}
{\renewcommand\arraystretch{1.2}
\small
\begin{tabular}{|c|c|c|c|c|c|}
\hline
$I^G (J^{PC})$ & threshold &  \multicolumn{2}{c|}{$E$ [MeV]} & \multicolumn{2}{c|}{decay channels} \\
\cline{3-6}
 & & $\pi$-potential & $\pi \rho\, \omega$-potential & s-wave & p-wave \\
\hline
$0^- (0^{--})$ & $\mathrm{B}\bar{\mathrm{B}}^{\ast}$ & no & no
	     &$\chi_{\mathrm{b}1} \!+\! \omega$ &  $\eta_{\mathrm{b}} \!+\! \omega$, $\Upsilon \!+\! \eta$ \\
\hline
$0^+ (1^{-+})$ & $\mathrm{B}\bar{\mathrm{B}}^{\ast}$ & $28.6-i91.6/2$ & $17.8-i30.1/2$
	     &$\mathrm{h}_{\mathrm{b}} \!+\! \omega^{\ast}$, $\chi_{\mathrm{b}1}
 \!+\! \eta$ & $\eta_{\mathrm{b}} \!+\! \eta$, $\Upsilon \!+\! \omega$ \\
\hline
$0^- (2^{+-})$ & $\mathrm{B}\bar{\mathrm{B}}^{\ast}$ & no & no  & --- &
 $\mathrm{h}_{\mathrm{b}} \!+\! \eta$, $\chi_{\mathrm{b}0,1,2} \!+\! \omega$ \\
\hline
\end{tabular}
}
\end{center}
\label{tbl:result_table_2}
\end{table}%

\section{Summary}
In this paper, we have systematically studied the possibility of the the $\mathrm{P}^{(\ast)}\bar{\mathrm{P}}^{(\ast)}$ bound and resonant states having exotic quantum numbers $I^G(J^{PC})$.
These states are consisted of at least four quarks, because their quantum numbers cannot be assigned by the quarkonium picture and hence they are genuinely exotic states.
We have constructed the potential of the 
 $\mathrm{P}^{(\ast)}\bar{\mathrm{P}}^{(\ast)}$ states using the effective
Lagrangian respecting the heavy quark symmetry.
The channel mixing due to the mass degeneracy of $\mathrm{P}$ and $\mathrm{P}^{\ast}$
plays an important role to form the $\mathrm{P}^{(\ast)}\bar{\mathrm{P}}^{(\ast)}$ bound and/or
resonant states.
We have numerically solved the Schr\"odinger equation with
the channel-couplings for the $\mathrm{P}^{(\ast)} \bar{\mathrm{P}}^{(\ast)}$ states with $I^{G}(J^{PC})$ for $J \le 2$.

As a result, in $I=1$, we have found that the
$I^{G}(J^{PC})=1^{+}(1^{+-})$ states have a bound state with binding 
energy 8.5 MeV, and a resonant state with the resonance energy 50.4 MeV 
and the decay width 15.1 MeV.
We have successfully reproduced the positions of
$\mathrm{Z}_{\mathrm b}(10610)$ and $\mathrm{Z}_{\mathrm b}(10650)$ observed by Belle.
Therefore, the twin resonances of $\mathrm{Z}_{\mathrm b}$'s can be interpreted as
the $\mathrm{B}^{({\ast})}\bar{\mathrm{B}}^{({\ast})}$ molecular type states.
It should be noted that the $\mathrm{B} \bar{\mathrm{B}}^{\ast}$-$\mathrm{B}^{\ast} \bar{\mathrm{B}}$, $\mathrm{B} \bar{\mathrm{B}}^{\ast}$-$\mathrm{B}^{\ast} \bar{\mathrm{B}}^{\ast}$ and $\mathrm{B}^{\ast} \bar{\mathrm{B}}$-$\mathrm{B}^{\ast} \bar{\mathrm{B}}^{\ast}$ mixing effects are important, because many structures disappear without the mixing effects.
We have obtained the other possible
$\mathrm{B}^{({\ast})}\bar{\mathrm{B}}^{({\ast})}$ states in $I=1$.
We have found  one bound state in  each $1^{+}(0^{--})$ and  $1^{-}(1^{++})$, one resonant state in $1^{-}(2^{++})$ and twin resonant states in each $1^{+}(1^{--})$ and $1^{+}(2^{--})$.
It is remarkable that another two twin resonances can exist in addition to the $\mathrm{Z}_{\mathrm{b}}$'s.
We have also studied the $\mathrm{B}^{({\ast})}\bar{\mathrm{B}}^{({\ast})}$ states in $I=0$ and found one resonant state in $0^{+}(1^{-+})$.
We have checked the differences between the results from the $\pi$
exchange potential and those from the $\pi \rho\, \omega$ potential, and found that the difference is small.
Therefore, the one pion exchange potential dominates as the interaction in the $\mathrm{B}^{(\ast)}\bar{\mathrm{B}}^{(\ast)}$ bound and resonant states.
We also study $\mathrm{D}^{(\ast)} \bar{\mathrm{D}}^{(\ast)}$ molecular states, but there are no
bound/resonant states.
Because the reduced mass is smaller and the mass splitting between $\mathrm{D}$ and $\mathrm{D}^{\ast}$ is larger compared with bottom sector.

The $\Upsilon(5S)$ decay is a useful tool to search the 
$\mathrm{B}^{({\ast})}\bar{\mathrm{B}}^{({\ast})}$ states.
$\Upsilon(5S)$ can decay to the
$\mathrm{B}^{(\ast)}\bar{\mathrm{B}}^{(\ast)}$ states with $1^{+}(0^{--})$, $1^{+}(1^{--})$ and $1^{+}(2^{--})$ by a single pion emission in p-wave and the state with $0^{+}(1^{-+})$ by $\omega$ emission in p-wave.
$\Upsilon(5S)$ can also decay to the
$\mathrm{B}^{(\ast)}\bar{\mathrm{B}}^{(\ast)}$ states 
with $1^{-}(0^{++})$, $1^{-}(1^{++})$ and $1^{-}(2^{++})$ by radiative decays.
In the future, various exotic states would be observed around the
thresholds from $\Upsilon(5S)$ decays in accelerator facilities such as Belle and also would be searched in the relativistic heavy ion collisions in RHIC and LHC~\cite{Cho:2010db,Cho:2011ew}.
If these states are fit in our predictions, they  will be good
candidates of the $\mathrm{B}^{({\ast})}\bar{\mathrm{B}}^{({\ast})}$ molecular states.

\Acknowledgements
We thank  Prof.~S.~Takeuchi and Prof.~M.~Takizawa for fruitful discussions and comments.
This work is supported in part by Grant-in-Aid for Scientific Research on 
Priority Areas ``Elucidation of New Hadrons with a Variety of Flavors 
(E01: 21105006)" (S.Y. and A.H.) and by ``Grant-in-Aid for Young Scientists (B)
22740174" (K.S.), from 
the ministry of Education, Culture, Sports, Science and Technology of
Japan.

\end{document}